\documentclass[aps,pl,amsmath,amssymb,twocolumn,letterpaper,%
         footinbib,bibnotes,showpacs,reprint,balancelastpage,raggedbottom,%longbibliography,%
         citeautoscript,floatfix,superscriptaddress]{revtex4-1}
\usepackage{graphicx}
\usepackage{bm}
\usepackage{dcolumn}
 \usepackage{xfrac}
\usepackage{xcolor}
\usepackage{microtype}
\usepackage[pdftex,
  breaklinks,             %
  bookmarks = false, %
  pdfpagemode   = UseNone,%
  pdfstartview 	= FitH,		%
  pdfstartpage  = 1,      %
  colorlinks    = true,   %
]{hyperref}

\begin{document}
\title{Magneto-electric coupling in type-I multiferroic ScFeO$_3$}

\author{G.\ Giovannetti}
\affiliation{CNR-IOM-Democritos National Simulation Centre}
\affiliation{International School
for Advanced Studies (SISSA), Via Bonomea 265, I-34136, Trieste, Italy}
\affiliation{Institute for Theoretical Solid State Physics, IFW-Dresden, PF 270116, 01171 Dresden, Germany}
\author{D.\  Puggioni}
  \affiliation{Department of  Materials  Science  and  Engineering,  Northwestern  University,  Evanston,  IL  60208,  USA}
  \author{P.\ Barone}
\affiliation{Consiglio Nazionale delle Ricerche (CNR-SPIN), Via Vetoio, I-67010 L'Aquila, Italy}
\affiliation{ Graphene Labs, Istituto Italiano di Tecnologia, via Morego 30, 16163 Genova, Italy}
\author{S.\ Picozzi}
\affiliation{Consiglio Nazionale delle Ricerche (CNR-SPIN), Via Vetoio, I-67010 L'Aquila, Italy}
\author{J.\ M.\  Rondinelli}
 \affiliation{Department of  Materials  Science  and  Engineering,  Northwestern  University,  Evanston,  IL  60208,  USA}
\author{M.\  Capone}
\affiliation{International School
for Advanced Studies (SISSA), Via Bonomea 265, I-34136, Trieste, Italy}
\affiliation{CNR-IOM-Democritos National Simulation Centre}

\begin{abstract}
%We investigate the electronic structure and the ferroelectric properties of the recently discovered multiferroic ScFeO$_3$ by means of density-functional theory and  dynamical mean-field theory. We characterize the material as an intermediately correlated compounds with a half-filled iron $3d$ manifold (Fe$^{3+}$).  This configuration naturally favors antiferromagnetic ordering and the intermediate interaction strength maximizes the N\'eel temperature, leading to a theoretical estimate in good agreement with the large experimental values.
%We find an exceedingly large ferroelectric polarization, $\sim$105\,$\mu$C/cm$^{2}$ related to the off-centering of Sc atoms but strongly sensitive to the local magnetization at the Fe atoms.
%This results in a strong negative magnetoelectric coupling, which is unexpected in type-I multiferroic materials, due to the different origin of the magnetic and ferroelectric orderings. 

We investigate the electronic structure and the ferroelectric properties of the recently discovered multiferroic ScFeO$_3$ by means of ab-initio calculations. The $3d$ manifold of Fe in the half-filled configuration naturally favors an 
%intermediate correlated regime and 
antiferromagnetic ordering, with a  theoretical estimate of the antiferromagnetic N\'eel temperature  in good agreement with the experimental values.
%, and responds to the electronic correlations.
We find that the inversion symmetry-breaking is driven by the off-centering of Sc atoms, which results in a large ferroelectric polarization of $\sim$105\,$\mu$C/cm$^{2}$. Surprisingly the ferroelectric polarization is sensitive to the local magnetization of the Fe atoms resulting in a large negative magnetoelectric interaction. This behavior is unexpected in type-I multiferroic materials because the magnetic and ferroelectric orders are of different origins. 

\end{abstract}

\pacs{}
\maketitle

\section{Introduction} Multiferroic (MF) materials display two or more ferroic orders, e.g., magnetism and ferroelectricity, often with a mutual interplay between the primary order parameters (magnetoelectric coupling) \cite{Hill}. 
Designing these materials to have high ferroic ordering temperatures and strong responses would allow them to serve as 
a materials platform for many practical logic and memory applications \cite{Cheong}. 

Multiferroics can be classified according to the origin of the coexisting 
magnetic and ferroelectric orders \cite{Khomskii}. 
In type-I multiferroics, ferroelectricity and magnetic order arise 
largely independent of each other, as each originates from different atomic sublattices; nonetheless, 
coupling between order parameters remains possible. 
Examples of type-I multiferroics are  BiFeO$_3$ \cite{Wang} and Sr$_{1-x}$Ba$_{x}$MnO$_3$ \cite{Sakai}, which present high magnetic ordering temperatures and large ferroelectric polarizations. 
% due to polar displacements which are unrelated to the magnetic ions\cite{NeatonBFO,GGSBMO}. 
%
Instead in type-II MFs, magnetism induces ferroelectricity, implying a 
strong coupling between the two orders.
Known examples of type-II multiferroics are $R$Mn$_2$O$_5$ \cite{Hur} and $R$Mn$O_3$ \cite{Kimura} ($R$ = Rare Earth). In these materials the low-temperature magnetic structure lifts inversion symmetry \cite{Picozzi123,GG125} and gives rise to the ferroelectric distortion.

The distinction between the two MF classes is expected to be reflected 
in the strength of the magnetoelectric coupling. It should  be much stronger in type-II multiferroics compared to type-I.
There are, however, examples that challenge this notion.
In BiFeO$_3$, a small negative magnetoelectric coupling with a variation of ferroelectric polarization ($\sim$40\,$n$C/cm$^{2}$) has been observed at the onset of the magnetic ordering \cite{Lee}. Also, the cycloidal modulation of its antiferromagnetic phase has been ascribed to an inhomogeneous magnetoelectric coupling, which rotates the direction of the magnetization and is uniquely determined by the ferroelectric polarization \cite{Kadomtseva, Johnson}.
A larger effect was found in Sr$_{1-x}$Ba$_{x}$MnO$_3$ where there is a substantial decrease in the electric polarization ($\sim$13$\mu$C/cm$^{2}$) at the magnetic critical temperature \cite{Sakai,GGSBMO}. 
These findings motivate us to re-examine the established notion 
of the magnetoelectric coupling strength in type-I multiferroics.
% and identify the origin of
%its unanticipated large response.
%with the aim of identifying the origin of the surprising magnetoelectric coupling in type-I multiferroics.

Recently ScFeO$_3$ has been synthesized under 15\,GPa at a temperatures above 1100\,K \cite{Kawamoto} and 
a number of other polymorphs can be realized in thin films \cite{Itoh}.
The high-pressure phase exhibits a polar $R3c$ space group with highly distorted ScO$_6$ and FeO$_6$ octahedra \cite{Kawamoto}.
It  exhibits weak ferromagnetism, of potential interest for applications \cite{Zhao}, with an high magnetic ordering temperature of 545\,K owing to a canted 
G-type antiferromagnetic (AFM) ordering of the Fe$^{3+}$ atoms \cite{Kawamoto}.

\begingroup
\squeezetable
\begin{table*}[]
\begin{ruledtabular}
\centering
\caption{\label{tab:latpar}Crystallographic 
parameters for the rhombohedral $R3c$ structure of ScFeO$_3$ obtained from PBE and PBE$+U$ calculations. The lattice parameters are fixed to 
the experimental values: $a=b=5.197$\,\AA, $c=13.936$\,\AA, $\alpha=\beta=90^\circ$, $\gamma=120^\circ$ reported in Ref.~\onlinecite{Kawamoto}.} %and
\begin{tabular}{llcccccccccccc}%
& & \multicolumn{3}{c}{PBE} & \multicolumn{3}{c}{PBE$+U=3$\,eV}& \multicolumn{3}{c}{PBE$+U=6$\,eV}  & \multicolumn{3}{c}{Experiment (Ref.\ \onlinecite{{Kawamoto}})} \\
\cline{3-5}\cline{6-8}\cline{9-11}\cline{12-14}\\[-0.8em]
Atom    &       Wyck.\ Site     & $x$   & $y$   & $z$  & $x$   & $y$   & $z$ & $x$   & $y$   & $z$ & $x$   & $y$   & $z$ \\
\hline \\[-0.8em]
Sc           &       $6a$              & 0                      &       0          &       0		 & 0                      &       0          &       0       & 0                      &       0          &      0   & 0                      &       0          &      0  		\\
Fe           &       $6a$              & 2/3                      &       1/3         &        0.122     & 2/3                      &      1/3         &        0.122 	& 2/3                      &       1/3         &       0.123      & 2/3                      &       1/3         &       0.123  	  \\
O            &       $18b$            & 0.353                      &       -0.022         &        0.061     & 0.352                      &       -0.023         &       0.060    	 & 0.353                      &       -0.023         &       0.058	& 0.349                      &       -0.022         &       0.061  \\
\end{tabular}
\end{ruledtabular}
\end{table*}
\endgroup

Multiferroic  ScFeO$_3$ shares the same $R3c$ space group of LiNbO$_3$ and BiFeO$_3$ and displays a ``mixture" of their electronic and ferroelectric properties.
%%GG 
%%it is expected that the perovskites materials with the greatest tendency toward octahedral rotations (those with the smallest tolerance factor $t$, defined as the geometric measure of ionic size mismatch of A and B ions in ABO$_3$ compounds) also have the largest ferroelectric instabilities \cite{Benedek}.
%
In materials with very small tolerance factor, $t$ \cite{note_tol}, as ScFeO$_3$ ($t=0.83$) and  LiNbO$_3$ ($t=0.85$), the $a^{-}a^{-}a^{-}$ tilt pattern in Glazer notation \cite{Glazer} is  electrostatically and energetically unstable because the $A$-site is severely underbonded. To stabilize the structure and optimize the environment of the $A$-site, a ferroelectric distortion which involves the $A$-cation is needed \cite{Benedek}. 
This is different from BiFeO$_3$ where the origin of the ferroelectric distortion is the stereochemical activity of the 6$s^2$ lone-pair of the Bi$^{3+}$ cation \cite{NeatonBFO}.
On the other hand, the $B$-site of ScFeO$_3$ is magnetic as in BiFeO$_3$, in contrast with the non-magnetic Nb cation in LiNbO$_3$. Note that ScFeO$_3$ has a very high magnetic ordering temperature in common with other multiferroic materials as Sr$_{1-x}$Ba$_{x}$MnO$_3$\cite{GGSBMO}, BiFeO$_3$ and PbNiO$_3$ \cite{Hao}.
We thus can classify ScFeO$_3$ as a type-I multiferroic, in which both ferroelectric and magnetic order exist albeit are expected to be weakly coupled. % despite its magnetoelectric coupling remaining to be investigated.
%both on the theoretical and experimental sides.

%In this work we explore, by means of first-principle calculations, how the ferroelectric polarization is strongly intertwined with the magnetic ordering even in type-I multiferroics and we address how this surprising effect and the properties of multiferroic ScFeO$_3$ are related to electron-electron correlations connected with the iron degrees of freedom.
%We relate the presence of sizable electronic correlations to the high critical magnetic ordering temperature of ScFeO$_3$ and suggest a recipe to find multiferroics with high ordering temperature in half-filling correlated materials close to the itinerant-to-localized transition.
%with electronic properties in between the itinerant and localized regime.
Here we study, by means of first-principles calculations,  the electronic, magnetic, and ferroelectric properties of ScFeO$_3$. 
%
%We propose that the high magnetic ordering temperature relies on the intermediate electron-electron correlation strength in the half-filled $3d$ manifold of the Fe$^{3+}$ cation, which would place the system on the verge of a paramagnetic Mott transition. % thus enhancing its N\'eel temperature.
%
%Indeed, 
Our theoretical estimate of the N\'eel temperature is 635\,K in good agreement with the experimental observations \cite{Kawamoto}. 
Next we find a large ferroelectric polarization of $\sim$105\,$\mu$C/cm$^{2}$ and examine its dependence on the magnetic order. %temperature of 
%We investigate how the ferroelectric polarization is intertwined with the magnetic ordering in this material. 
%Furthermore, 
We find evidence of strong magnetoelectric coupling between the local Fe magnetization and the electronic contribution to the total electric polarization, which suggests that the amplitude rather than the direction of the magnetization in collinear magnets may be important in other type-I multiferroics. % and need further investigation both experimentally and theoretically.

%We suggest i) the high critical magnetic ordering temperature of ScFeO$_3$ to be related to the presence of sizable electronic correlations and ii) a recipe to find multiferroics with high ordering temperature and strong magneto-electric coupling in half-filling correlated materials close to the itinerant-to-localized transition.

\section{Calculation Details}
We perform spin-polarized density functional  calculations within the Perdew-Burke-Ernzerhof approximation (PBE) \cite{PBE} and the PBE+$U$ method \cite{DFTplusU} as 
implemented in the Vienna {\it Ab initio} Simulation Package
(VASP) \cite{VASP} with the projector augmented wave (PAW) method \cite{PAW} to treat the core and valence electrons using the following electronic configurations $3p^{6}4s^{2}3d^{1}$ (Sc), $4s^{2}3d^{6}$ (Fe), $2s^{2}2p^{4}$ (O).
A kinetic energy cutoff
energy of 400\,eV is used to expand the wave functions and a $\Gamma$
centered  8$\times$8$\times$4 $k$-point mesh combined with the tetrahedron and Gaussian methods is used for Brillouin zone integrations. The ions are relaxed toward equilibrium until the Hellmann-Feynman forces are less than 1\,meV\,\AA$^{-1}$ whereas the cell parameters are fixed to the experimental values \cite{Kawamoto}.

It is well known that the PBE often underestimates the size of the band gap in systems with strongly localized $d$ orbitals, therefore we also calculated the structural and electronic properties within the rotationally invariant PBE+$U$ method \cite{DFTplusU} which requires two parameters, the Hubbard parameter $U$ and the exchange interaction $J_H$. In this work, we  fix the value of the Hund's exchange energy to $J_H =0.9$\,eV, as proposed for BiFeO$_3$ \cite{Gonzalez-Vazquez,Shorikov} and vary the magnitude of the Hubbard parameter $U$ between 3 and 6\,eV for the Fe $3d$-states. Note that the standard spin-polarized PBE  corresponds to $U=J_H=0$\,eV. 
The electric polarization is calculated using the Berry's phase method \cite{Vanderbilt} with 6 $k$-points for each string along the $c$ direction. 

Classical Monte Carlo (MC) simulation, with 20$\times$20$\times$20 supercells and 10$^{7}$ MC steps, is used to evaluate the N\'eel temperature.
\section{Results and discussion}
\subsection{Structure}

In Table~\ref{tab:latpar} we report the atomic positions within the $R3c$ space group and the experimental lattice parameters \cite{Kawamoto} at the PBE and PBE+$U$ with $U$=3 and 6\,eV levels. Results for other values of U are not shown due to the similarity in results with the case $U$=3 and 6\,eV. The atomic coordinates are in close agreement with the experimental ones \cite{Kawamoto} and are slightly affected by the Hubbard correction. The structure exhibits $a^{-}a^{-}a^{-}$ tilt pattern and large displacements of the Sc atoms. In the experimental structure, the FeO$_6$ octahedron is distorted with three short and three long Fe--O bonds of 1.96\,\AA\, and 2.15\,\AA\, respectively, resulting in an interoctaheral  O--Fe--O bond angle of 135$^\circ$. In our calculation this local environment is well reproduced with PBE$+U$ with $U$=3\,eV. In the remainder of this paper, we report results obtained using the experimental structure without ionic relaxations owing to the small difference with the equilibrium structures obtained from DFT. %theoretical ones. 

%\cite{Kawamoto}. The polar structure, which is similar to LiNbO$_3$, is characterized by out-of-phase tilting of corner-shared FeO$_6$ octahedra  (Glazer notation a$^{-}$a$^{-}$a$^{-}$) about the three-fold pseudocubic [111] axis along which Sc cation displacements also occur. This results in a spontaneous polarization along the %[111] direction, which corresponds to the $c$ axis of the experimental setting \cite{Kawamoto}. 

\subsection{Electronic and Magnetic Properties}
\begin{figure}
\centering
\includegraphics[width=0.9\columnwidth,angle=0]{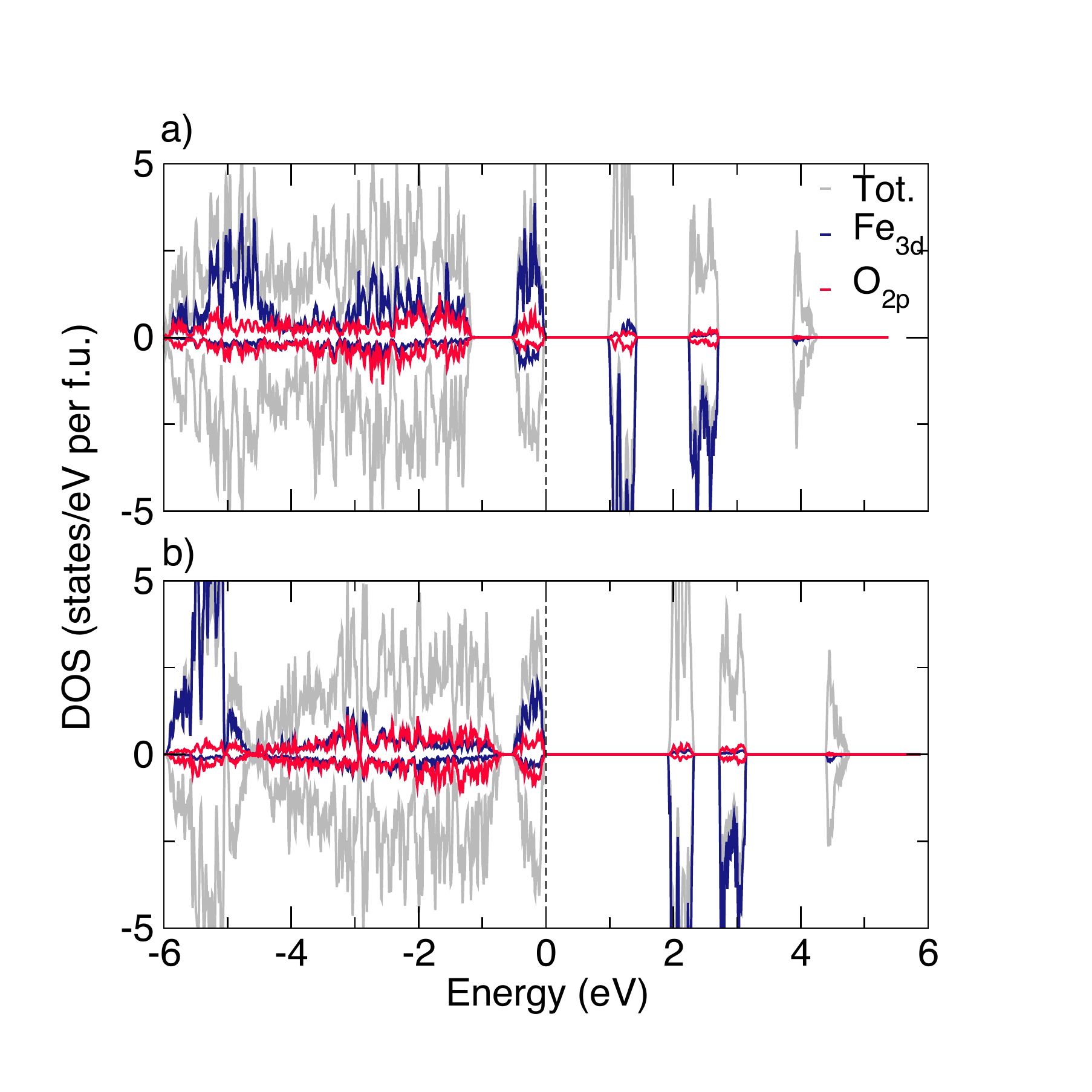} \\
%\vspace{0.1cm}
\caption{(Color online) 
The total (gray line) and  projected density of states of rhombohedral ($R3c$) ScFeO$_3$ with G-type AFM order calculated (a) with PBE and (b) PBE+$U$ ($U$=3.0, J$_H$=0.9 eV). The zero energy is set to the top of the valence band (dashed line). 
%Inset: Imaginary part of the self-energies $\Sigma$ within PBE+DMFT scheme for t$_{2g}$ and e$_g$ orbitals.
}
\label{fig1}
\end{figure}
%\emph{Role of Correlation on $T_N$.---}
Figure \ref{fig1}a shows the total and projected densities-of-states, calculated within the PBE for the $R3c$ structure of ScFeO$_3$. The ground state is insulating with a band gap of 0.96\,eV and exhibits G-type antiferromagnetic (AFM) order with a local magnetic moment of 3.7$\mu_B$ on the Fe atoms. The valence band is composed of Fe $3d$ states with $t_{2g}$ and $e_g$ orbital character strongly hybridized with O $2p$ states, consistent with a high-spin $d^5$ cation.  The strong hybridization between the Fe 3$d$ and O $2p$ valence electrons, in the energy range of -6 to 0\,eV, suggests that the Fe--O bonding is highly covalent. % at odds with the quite different electronegativity between Fe and O atoms. 
%
%Owing to the band filling and crystal field splitting the $e_g$ states are split from the $t_{2g}$ states by $\sim$0.6\,eV.
% can be divided into lower and upper parts.
%qui

The size of the band gap and the local magnetic moment increase as a function of $U$ within the PBE$+U$ formalism; in particular for $U=3$\,eV we find a band gap of 1.9\,eV (Figure \ref{fig1}b) and a local magnetic moment of 4$\mu_B$ on the Fe atoms in better agreement with the experimental property measurement data \cite{KojiFujita,Kawamoto}. The introduction of the Hubbard parameter also influences the ionicity of the Fe--O bonding. Indeed, the Fe occupied 3$d$ states are pushed down to lower energy, indicating that the Fe--O bonding is more ionic and the electrons are more localized on the atomic sites, while the Fe unoccupied 3$d$ states are pushed to higher energy.

To evaluate the magnetic ordering temperature we map the total energy of different magnetic phases on a Heisenberg model,  $H=-J\sum_{ij}\,S_i\cdot S_j$, describing classical spins interacting only with nearest neighbors. We follow the approach of Ref.~\cite{Lampis} and we calculate the nearest-neighbor exchange coupling $J$ of ScFeO$_3$ from the energy difference, calculated within PBE+$U$ with $U$=3\,eV, between the ferromagnetic (FM) and G-type AFM order \cite{NotaJ}. 
Assuming $S=5/2$ for Fe$^{3+}$ spins, we find an exchange interaction $J=$ -3.3\,meV, where the minus sign indicates the antiferromagnetic coupling. Using this magnetic coupling in classical Monte Carlo (MC) simulations we estimate a N\'eel temperature of 635\,K which compares reasonably well with the experimental value of 545\,K \cite{Kawamoto} given the strongly localized approximation of the adopted nearest-neighbor Heisenberg model.

\subsection{Ferroelectricity} 
\begin{figure}%[ht]
\centering
\includegraphics[width=0.99\columnwidth,angle=0]{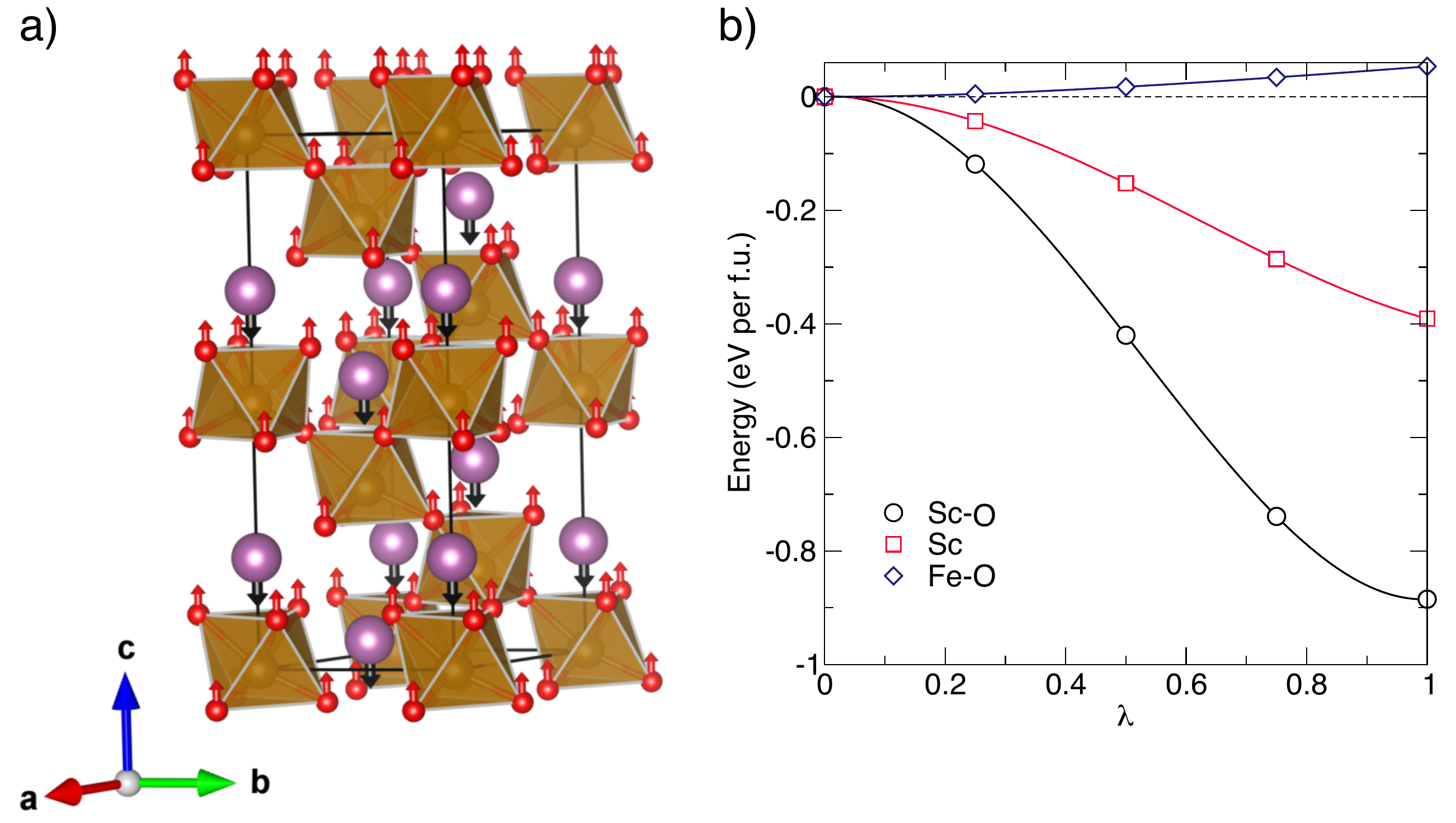} \\
%\vspace{0.2cm}
\caption{(Color online) (a) Illustration of the ferroelectric displacements by the Sc and O atoms. (b) Energy gain ($\Delta$E) as a function of the polar displacements ($\lambda$) within the PBE$+U$ ($U=3.0$\,eV, $J_H=0.9$\,eV) with respect to the centrosymmetric $R\bar3c$ structure. We compare the energy of the Sc-O mode with partial modes (Sc,Fe-O) where only the listed atoms are displaced.}
\label{fig2}
\end{figure}

Figure~\ref{fig2}a depicts the polar displacements of the Sc and O ions in the $R3c$ crystal structure. 
These displacements are along the [111] pseudocubic direction.  This results in a spontaneous polarization along the %[111] direction, which corresponds to the 
$c$ axis in the hexagonal setting of rhombohedral structure.
The geometry-induced inversion symmetry-breaking of the $R3c$ structure with respect to the centrosymmetric $R\bar{3}c$ phase can be described by a polar mode with A$_{2u}$ symmetry. Along the polar  $c$ axis the Sc and O ions have opposite displacements of -0.55 and 0.29\,\AA, respectively, while the Fe cations under go minor displacements and contribute weakly to the electric polarization (Fig.\ \ref{fig2}a). 
The situation along the non-polar  $a$ and $b$ axes is different; the Sc and O ions exhibit antipolar displacements. %that do not permit an electric polarization.

Using PBE+$U=3$\,eV and the G-type magnetic order, we compute the energy gain ($\Delta$E) of the  polar $R3c$ structure with respect to the centrosymmetric $R\bar3c$ structure as a function of $\lambda$, which  is a dimensionless parameter that continuously connects the centrosymmetric $R\bar3c$ structure ($\lambda$=0) to the experimentally-determined polar 
$R3c$ structure ($\lambda$=1).
% as $\Delta E = E(\lambda) - E(\lambda=0)$.
% a parameter $\lambda$ continuously connecting the centrosymmetric ($\lambda = 0$) structure to the experimentally-refined polar  state ($\lambda = 1$) as $\Delta E = E(\lambda) - E(\lambda=0)$.
%
Fig.\ \ref{fig2}b  shows the change in energy obtained by displacing the Sc and O (Sc-O) ions participating in the $A_{2u}$ mode versus the contributions owing to the displacements of two other subsets of ions, namely the Sc cation only and the Fe and O (Fe-O) ions. We mention in this context that the ionic positions of $R3c$ and $R\bar3c$ structures are interpolated by a linear relation, which is a standard procedure in the study of many multiferroic and ferroelectric materials \cite{AKEN,Puggioni,Cohen}.
%, whereby the Sc displacements are neglected. 
%
%The results for the total energy gain $\Delta E = E(\lambda) - E(\lambda=0)$ reported in Fig. \ref{fig2} show
We find  that the $R3c$ structure is stabilized by either displacing the Sc cation or the Sc-O ions and that Fe-O distortions alone are unfavorable, leading to an increase of the total energy. 
Consistent with the small tolerance factor argument \cite{Kawamoto,Benedek} and our finding that the largest contribution to the stabilization of the polar structure comes from the  polar Sc and O displacements, we conclude that the ferroelectric phase arises as a consequence of the Sc cation size. Note that the large energy-gain difference between that obtained from the Sc-O mode and the Sc mode demonstrates the decisive role played by the oxide anion. % in the  stability of the structure.% and the coupling between the different ions.
\begin{figure}%[ht]
\centering
\includegraphics[width=0.9\columnwidth,angle=0]{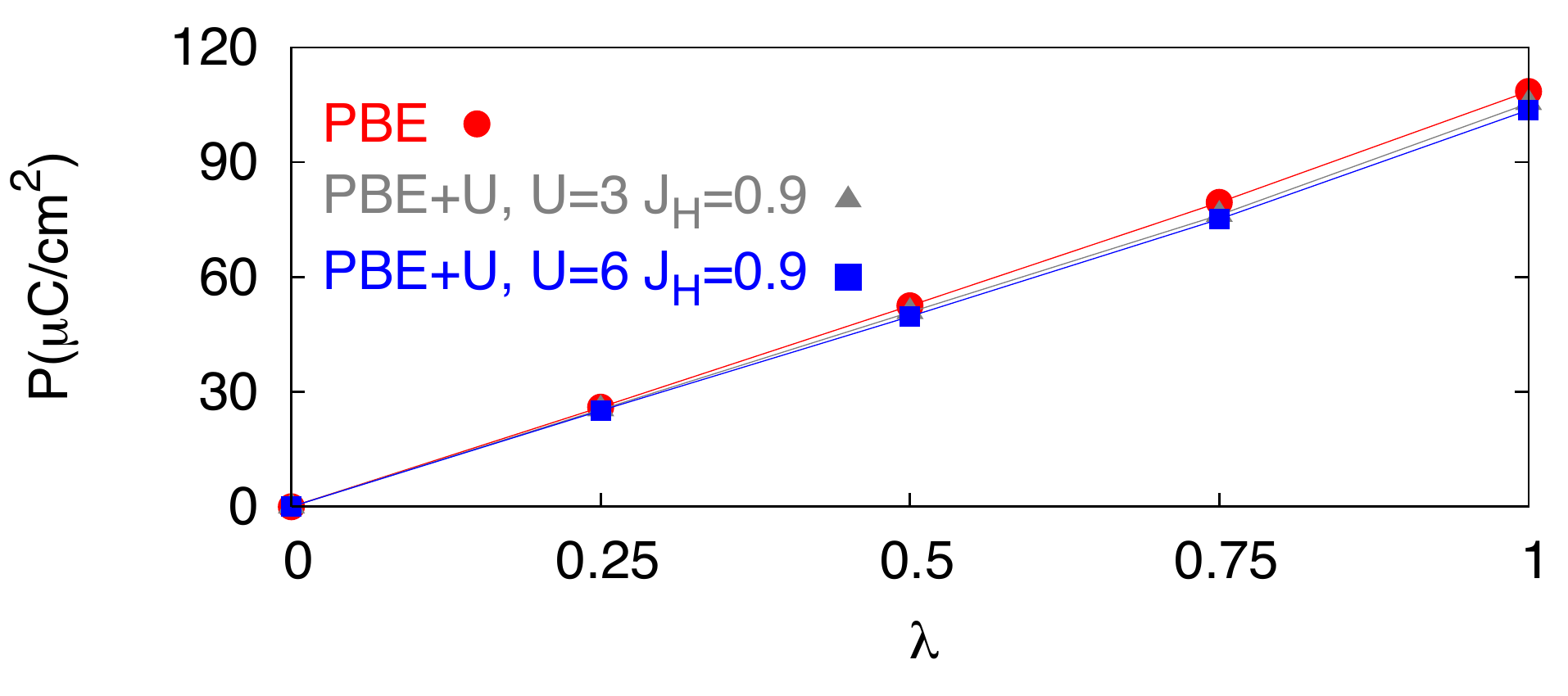} \vspace{-\baselineskip}
%\vspace{0.5cm}
\caption{(Color online) (a) The functional dependent electric polarization ($P$) for ScFeO$_3$ along the path connecting the paraelectric ($\lambda = 0$) and the ferroelectric structures ($\lambda = 1$). %(b) Ionic (I) and electronic (E) contributions to the ferroelectric polarization of different ions (Sc,O) contributing to the ferroelectric polarization as function of $\lambda$ for $U$=3\,eV and J$_H$=0.9\,eV.
}
\label{fig3}
\end{figure}

\begingroup
%\squeezetable
\begin{table}
\begin{ruledtabular}
\centering
\caption{\label{tab:bec}Born effective charges ($Z^*$) for displacements
along  the [111]-pseudocubic direction  for ScFeO$_3$ in the experimental $R3c$ crystal structure  obtained from PBE and PBE$+U$ calculations. Nominal oxidation states specified in parentheses.} %and
\begin{tabular}{lccc}%
%\multicolumn{4}{l}{$Pnma$ (62)} $a=5.5909$\,\AA, $b=7.8821$\,\AA, $c=5.5617$\,\AA \\
                            &       Sc  ($+3$)  & Fe   ($+3$) & O   ($-2$) \\
\hline\\[-0.8em]
PBE                    &       3.815                & 3.688                            &      -2.505     \\
PBE$+U$=3\,eV          &      3.798                & 3.650                            &      -2.477        \\
PBE$+U$=6\,eV         &       3.780                & 3.504                            &      -2.433             \\
\end{tabular}
\end{ruledtabular}
\end{table}
\endgroup

Next, we evaluate  the ferroelectric polarization using the Berry phase approach \cite{Vanderbilt}.
In Fig.~\ref{fig3} we show the total ferroelectric polarization $P$ as a function of the Sc-O displacement mode. %which we indicate with $\lambda$. As one can see, 
We find that $P$ is weakly sensitive to the Hubbard parameter as expected by the small influence $U$ has on the crystal structure (Table~\ref{tab:latpar}): 
The magnitude of the electric polarization is 
108.5\,$\mu$C/cm$^{2}$, 105.5\,$\mu$C/cm$^{2}$, and 103.6\,$\mu$C/cm$^{2}$ for 
$U=J_H=0$\,eV, $U=3.0$\,eV, and $U=6.0$\,eV, respectively. 

In Table~\ref{tab:bec} we show the Born effective charges ($Z^*$) for the $R3c$ structure of ScFeO$_3$ as a function of the different levels of theory used in this work. In agreement with Fig.~\ref{fig3}, we find that $Z^*$ decreases slightly as a function of correlation $U$. Indeed the polarization can be written as $P \propto \sum_i Z_i ds_i$, where $ds_i$ are the ferroelectric displacements and $Z_i$ are the Born effective charges. Increasing $U$ while keeping the ferroelectric displacements fixed ($ds_i$) we find that the effective charges get closer to the formal nominal values Sc$^{3+}$, Fe$^{3+}$ and O$^{2-}$, which implies a smaller anomalous (electronic) contribution to $P$. Also, the $Z^*$ of Sc is somewhat larger than the nominal oxidation state, indicating the importance of this ion in the the polar distortion. This is similar to BiFeO$_3$ \cite{NeatonBFO} but different from LiNbO$_3$ \cite{Veithen:2002},  NaNbO$_3$, and  KNbO$_3$ \cite{Zhong:1994} where the $Z^*$ of the $A$ cations are much closer to the nominal values.

\begin{figure}[t]
\centering
\includegraphics[width=0.9\columnwidth,angle=0]{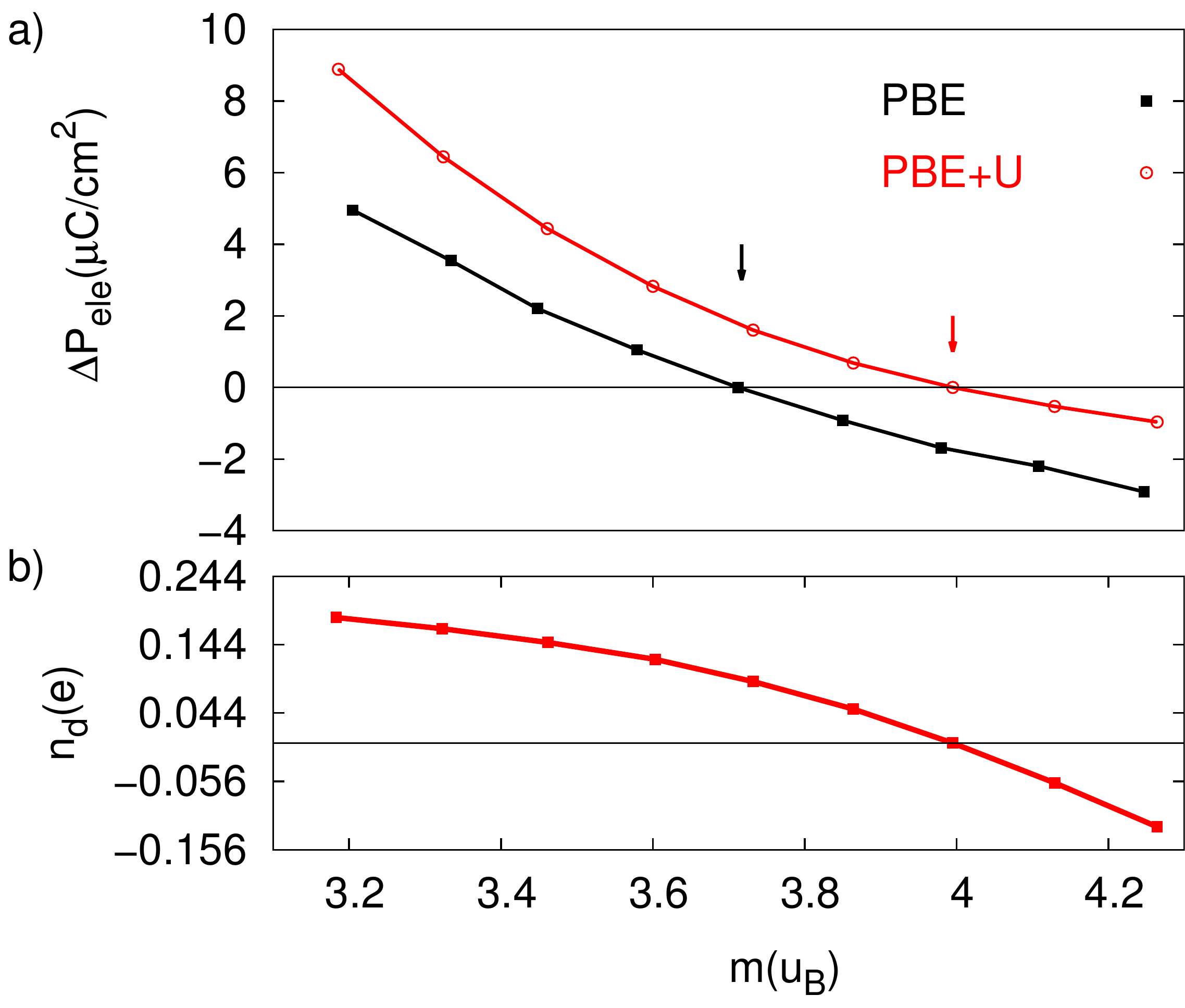} \vspace{-1\baselineskip}
%\vspace{0.5cm}
\caption{(Color online)  a) Change of the electronic contribution ($\Delta$P$_{e}$) to ferroelectric polarization for ScFeO$_3$ within PBE and PBE+U (U=3.0\,eV, J$_H$=0.9\,eV) as function of the local Fe magnetization at fixed experimental ionic positions. The zero of P$_{ele}$ is set to the value of the self-consistent calculations without constraining local Fe magnetization. b) Change of Fe charge as function of the local Fe magnetization at fixed experimental ionic positions.}
% Arrows indicates the values of local magnetization $m$ at Fe sites at self-consistent PBE and PBE+U calculations.}
\label{fig4}
\end{figure}

\subsection{Magnetoelectric Coupling} 

\subsubsection{Pure Electronic contribution to P}

% As we have shown above the Hubbard parameter $U$ controls both the covalency of the crystal and the the local magnetization at Fe sites. Therefore we expect a strong coupling between ferroelectric polarization and local magnetization in ScFeO$_3$.

Although the ferroelectric polarization is mainly driven by the Sc and O atoms, we now investigate the possible existence of a magnetoelectric coupling.
% as in BiFeO$_3$ \cite{Lee}. 
We use the experimental $R3c$ crystal structure %\cite{Kawamoto},  we 
to evaluate the \emph{electronic} contribution to the ferroelectric polarization, which is a sum of the electronic and ionic contributions, by performing constrained magnetic calculations whereby 
the amplitude of the local Fe magnetic moment is varied (Fig.~\ref{fig4}a).
This change in the electronic polarization as a function of the local magnetic moment on the Fe site is the order of the 1\,$\mu$C/cm$^{2}$ and  larger than that observed in BiFeO$_3$ ($\sim$ 40$n$C/cm$^{2}$) \cite{Lee}.
%between 3.2\,$\mu_B$ and 4.2\,$\mu_B$. 
%using the experimental $R3c$ crystal structure \cite{GG1111}.
%%%%%%
%%%%%%
%To make this link explicit, we evaluate the electronic contribution to the ferroelectric polarization performing constraining in calculations where the amplitude of the local Fe magnetic moments is constrained at different values using the experimental $R3c$ crystal structure \cite{GG1111}. 
%
%Our calculations, regardless of the approach used, confirm the hypothesis, as shown in Fig.~\ref{fig4}. Indeed, 
We find that the amplitude of the local Fe moment in the ordered G-AFM state controls the electronic $P_e$ contribution to $P$. 

The coupling between the electronic polarization and the local magnetic moment can be understood as a consequence of the decrease/increase in the amount of static electronic charge at the Fe site, which is donated from the oxide ions active in the inversion symmetry-breaking displacements. Indeed by inspecting the on-site density matrix of Fe ions and summing up over its eigenvalues we find that increasing the value of local magnetic moment a given amount of electronic charge is moved from Fe to O ions (see Fig.~\ref{fig4}b).
%of ScFeO$_3$.
%is indeed a consequence of the same physics we invoked to understand the decrease of polarization as a function of the Hubbard U. 
%We verified this explicitly by computing the Born effective charge as a function of the magnetic moment at a fixed value of the $U$ parameter. Increasing the magnetic moment pushes indeed the values of Z$_{Sc}$, Z$_{Fe}$, Z$_{O}$ closer to the nominal values, reproducing the qualitative effect of increasing the interaction U. % 
%MC is this what you meant?
%We find in our calculations that the effect of the changes of the local Fe magnetization is to decrease/increase the amount of static electronic charge at Fe/O ions. 
%To further substain this point of view we calculate the Born effective charge as function of the  size of the local Fe magnetic moments and we find that increasing local Fe magnetic moments push Z$_{Sc}$, Z$_{Fe}$, Z$_{O}$ closer to the nominal values.
%These results highlight how, once the polar displacements are active, electronic correlations determine the value of the magnetization and this dictates the electronic contribution to the total ferroelectric polarization by constraining the d/p orbital Fe/O occupancies. 
These results highlight how, once the polar displacements are active, the local spin magnetization influences the electronic contribution to the total ferroelectric polarization by changes in the $d/p$ orbital Fe/O occupancies.

This coupling between magnetism and the electric polarization results in a `negative' magnetoelectric coupling in ScFeO$_3$; `negative' in the sense that the magnetic order suppresses the ferroelectric polarization. 
%
% Experimentally the changes of the ferroelectric polarization related to the onset of the magnetic ordering should be observed close to the magnetic ordering temperature of ScFeO$_3$ monitoring the ferroelectric polarization as a function of temperature. 
We conjecture that the same hidden magnetic effect identified in ScFeO$_3$ is likely common in other multiferroic materials like, e.g., BiMnO$_3$ \cite{KimuraBiMnO3} and Sr$_{1-x}$Ba$_{x}$MnO$_3$ \cite{Sakai}.
Last, we remark here that we did not investigate the role of canting in the AFM structure (and the consequent weak ferromagnetism), but our conclusions on the local magnetization-dependent electric polarization are not expected to be sensitive to the details of the AFM state.
%, as they are based on generic effects of correlations.
%

\subsubsection{Ionic and Electronic contribution to P}
\begin{figure}
\includegraphics[width=0.92\columnwidth]{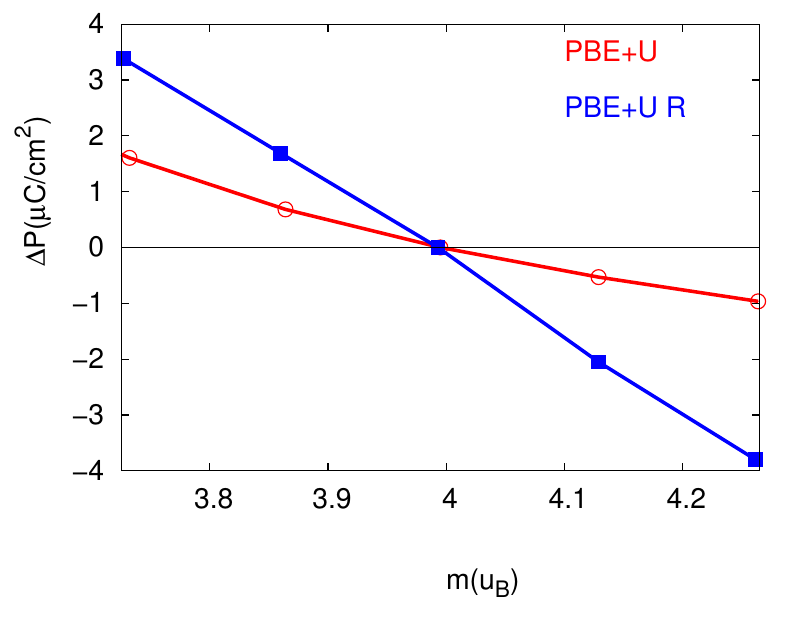} \\
\caption{(Color online) Change of the ferroelectric polarization ($\Delta$P) for ScFeO$_3$ within PBE+U (U=3.0\,eV, J$_H$=0.9\,eV) as function of the local Fe magnetization without (PBE+U) and with ionic relaxation (PBE+U R). The zero of P is set to the value of the self-consistent calculations without constraining local Fe magnetization.}
\label{fig5}
\end{figure}
As further step, in the understanding of the relation between ferroelectric polarization and onset of the magnetic ordering, we allow the ions to relax by keeping the magnetic moment at a fixed value and then we recalculate the ferroelectric polarization as sum of electronic and ionic contributions. As shown in Fig.~\ref{fig5}, we find that the magneto-electric coupling is further enhanced by inclusion of ionic relaxations. This clearly suggests  that the ferroelectric displacements are controlled by the local magnetization of Fe sites. Note that increasing the local magnetization of the Fe sites, the contribution to the ferroelectric polarization of the Sc and O ions is only slightly decreased.

\section{Conclusions} 
%Using a variety of first-principles methods including interaction effects (PBE, PBE+$U$, PBE+DMFT), we investigate the electronic structure, the role of electronic correlations and the origin of the ferroelectric state in ScFeO$_3$. We find that electronic correlations are responsible for the ordering of the iron magnetic moments while the ferroelectric instability is mainly driven by coupled ferroelectric distortions of Sc and O sites. The high magnetic ordering temperature descends from an intermediate degree of correlation which maximizes the N\'eel temperature. 
Using first-principles methods that include interaction effects (PBE, PBE$+U$), we investigated the electronic and magnetic structure and the origin of the ferroelectric state of ScFeO$_3$. We find that the ferroelectric instability is mainly driven by coupled polar displacements of Sc and O  similar to proper ferroelectrics such as LiNbO$_3$. 
%The G-type antiferromagnetic structure originates  from an intermediate degree of electron correlation and 
The electric polarization is found to be sensitive to the magnitude of the local magnetic moment on the Fe sites.
%, is found to originate related to a mixture of itinerant and localized electrons, combined with the inversion symmetry breaking stabilizes a multiferroic ground state.  
%The high magnetic ordering temperature descends from an intermediate degree of correlation which maximizes the N\'eel temperature.
%
%Besides determining the magnetic ordering, strong correlations induce significant changes in the covalency of the crystal, which reflect in the value of the electronic polarization, which decreases when the interaction increases.  This leads to a sizable negative magneto-electric coupling in ScFeO$_3$, which opens new perspectives  in the field of multiferroic materials driving their search in compounds close to Mott instability. 
We find a `negative' magnetoelectric interaction in type-I multiferroic ScFeO$_3$, which may also be active in other multiferroic compounds. %driving their search in compounds close to Mott instability.

\begin{acknowledgments}
G. G.\ acknowledges Koji Fujita for sharing experimental data and discussions.
G. G.\ and M. C.\ acknowledge financial support by European Research Council
under FP7/ERC Starting Independent Research Grant ``SUPERBAD" (Grant Agreement n. 240524). 
D.P.\ and J.M.R.\ were supported by the ARO under grant no.\ W911NF-15-1-0017 for financial support.
\end{acknowledgments}

%Unused bibitems


\begin{thebibliography}{99}
\bibitem{note_tol} The tolerance factor, $t$, can be used as a measure of the degree of distortion of a ABO$_3$ perovskite from ideal cubic structure. It is defined as $t=\frac{r_A+r_0}{\sqrt2(r_B+r_0)}$, where $r_A$ is the radius of the $A$-cation, $r_B$ is the radius of the $B$-cation, and $r_0$ is the radius of the anion.
\bibitem{NotaJ} The total energy of the FM-type and G-type can be written as: \\
E$_{FM-type}$ = E$_{NM}$ - 6 $J$ S$_{i}$ S$_{j}$ \\
E$_{G-type}$ = E$_{NM}$ + 6 $J$ S$_{i}$ S$_{j}$ \\
in which E$_{NM}$ represent the total energy of non-mangetic ground state (we assume S = $\frac{5}{2}$).
\end{thebibliography}
\end{document}